\documentclass[11pt]{article}
\usepackage{epsf} 
\usepackage{amsmath}
\usepackage{amssymb}
\usepackage{epsfig}
\usepackage{latexsym}
\usepackage{amsfonts}
\usepackage{graphicx}
\begin{document}
\parindent 0mm 
\setlength{\parskip}{\baselineskip} 
\thispagestyle{empty}
\pagenumbering{arabic} 
\setcounter{page}{1}
\mbox{ }
\hfill UCT-TP-303/14, MITP/14-093

\hfill Revised: June 2015

\begin{center}
{\Large {\bf Determination of the gluon condensate from data in the charm-quark region}}\\
\end{center}
\vspace{.05cm}

\begin{center}
{\bf  C. A. Dominguez}$^{(a)}$, 
{\bf L. A. Hernandez}$^{(a)}$, 
{\bf K. Schilcher}$^{(a),(b)}$, \\ 
\end{center}

\begin{center}
{\it $^{(a)}$Centre for Theoretical and Mathematical Physics, 
and Department of Physics, University of
Cape Town, Rondebosch 7700, South Africa}
\\

{\it $^{(b)}$PRISMA Cluster of Excellence, Institut f\"{u}r Physik, 
Johannes Gutenberg-Universit\"{a}t, D-55099 Mainz, Germany}
\\
\end{center}

\begin{center}
\footnotesize
{\it E-mail:} 
cesareo.dominguez@uct.ac.za, 
HRNLUI001@myuct.ac.za, 
\\
karl.schilcher@uni-mainz.de, 

\end{center}

\begin{center}
\textbf{Abstract}
\end{center}
The gluon condensate, $\langle \frac{\alpha_s}{\pi} G^2 \rangle$, i.e. the leading order power correction in the operator product expansion of current correlators in QCD at short distances, is determined from $e^+ e^-$ annihilation data in the charm-quark region. This determination is based on finite energy QCD sum rules, weighted by a suitable integration kernel to (i) account for potential quark-hadron duality violations, (ii)  enhance the contribution of the well known first two narrow resonances, the $J/\psi$ and the $\psi(2S)$, while quenching substantially the data region beyond, and (iii) reinforce the role of the gluon condensate in the sum rules. By using a kernel exhibiting a singularity at the origin, the gluon condensate enters the Cauchy residue at the pole through the low energy QCD expansion of the vector current correlator. These features allow for a  reasonably precise determination of the condensate, i.e. $\langle  \frac{\alpha_s}{\pi} G^2 \rangle =0.037 \,\pm\, 0.015 \;{\mbox{GeV}}^4$.\\
  

ArXiv ePrint: 1411.4500
  
\clearpage

\noindent
\section{Introduction}
One of the two-pillars of QCD sum rules (QCDSR) \cite{QCDSR1}-\cite{QCDSR2}, an analytic method to obtain results in QCD, is the operator product expansion (OPE) of current correlators at short distances beyond perturbation theory, to wit. Given a current-current correlation function of the squared four-momentum, $\Pi(Q^2)$, the OPE is formally written as
\begin{equation}
\Pi(Q^2)|_{QCD} = C_0\,\hat{I} + \sum_{N=1} \frac{ C_{2N}(Q^2,\mu^2)}{Q^{2N}}\;\langle {O}_{2N}(\mu^2)\rangle ,
\end{equation}
where $\langle {O}_{2N}(\mu^2)\rangle$ is short for $\langle 0| {O}_{2N}(\mu^2)|0 \rangle$, $\mu^2$ is a renormalization scale, $Q^2 < 0$ is the squared four-momentum, and the Wilson coefficients in this expansion, $ C_{2N}(Q^2,\mu^2)$,  depend on the Lorentz indices and quantum numbers of the current $J(x)$ entering the correlator, and  of the local gauge invariant operators $O_{2N} (\mu^2)$ built from the QCD quark and gluon fields. These operators are ordered by increasing dimensionality and the Wilson coefficients, calculable in perturbative QCD (PQCD), fall off by corresponding powers of $Q^2$ (explicitly factored out in Eq.(1)). In other words, this OPE achieves a factorization of short distance effects encapsulated in the Wilson coefficients, and long distance dynamics present in the vacuum condensates.
Since there are no gauge invariant operators of dimension $d=2$ involving the quark and gluon fields in QCD, it is normally assumed that the OPE starts at dimension $d=4$. This is supported by contemporary results from QCDSR analyses of $\tau$-lepton decay data \cite{C2a}-\cite{GG}, and $e^+ e^-$ annihilation data in the light-quark sector \cite{C2d}, which show no evidence for $d=2$ operators. A similar result is also found in lattice QCD (LQCD) analyses \cite{d2LQCD1}-\cite{d2LQCD2}. With the exception of the quark condensate, the numerical values of the vacuum condensates cannot be calculated analytically from first principles as this would be tantamount to solving QCD exactly. They can be determined e.g. from numerical simulations in lattice LQCD, or by confronting the OPE with suitable experimental data, as described in the sequel. In the chiral limit the first non-vanishing power term in the OPE with dimension $d=4$ has been traditionally identified with the gluon condensate \cite{QCDSR1}-\cite{QCDSR2}, \cite{ZAKHA}, $C_4 \langle{\cal{O}}_4\rangle = \frac{\pi}{3} \langle\alpha_s G^{a}_{\mu\nu} G^{a \mu \nu}\rangle$. Having the lowest dimension it dominates the OPE and thus QCDSR analyses of chirality conserving amplitudes, such as e.g. the Adler function. This condensate is also directly related to the vacuum energy density, $\epsilon$, through

\begin{equation}
\epsilon = \frac{\pi}{8 \alpha_s^2} \; \beta(\alpha_s) \; \langle\frac{\alpha_s}{\pi} G^a_{\mu\nu} G^{a \mu \nu}\rangle \;,
\end{equation}

where $\beta(\alpha_s)$ is the Gell-Mann-Low beta-function normalized as $\beta_1 = -\frac{1}{2}\left(11 - \frac{2}{3} n_F\right)$. The sign and the magnitude of the gluon condensate are of fundamental importance in the understanding of the strong interactions. A negative value of the vacuum energy $\epsilon$ is expected from models such as the bag model  and the instanton gas model. In addition, the numerical value of the gluon condensate should be chiral symmetric, i.e. determinations from a vector channel correlator should give the same value as those from an axial-vector channel correlator. In spite of more than 35 years of efforts to determine this condensate there is still  no clear consensus on its numerical value. There are at least three approaches to determine the gluon condensate. A direct, numerical approach consists in computing the average plaquette in LQCD. Unfortunately, an important and large perturbative component needs to be subtracted in this approach \cite{d2LQCD1}, and numerical results cover a huge range \cite{d2LQCD1}, \cite{LATTa}-\cite{LATTb}.
The other two approaches to determine the power corrections in the OPE are based on QCDSR, whose second pillar is the so-called quark-hadron duality. This is based on the use of the complex squared energy $s$-plane  to invoke Cauchy's theorem to relate QCD with the hadronic sector. Stable hadronic states enter as poles in the current correlator on the real $s$-axis, and resonances as singularities in the second Riemann sheet. These singularities lead to a discontinuity across the positive real  $s$-axis. Choosing a circular integration contour, and given that there are no other singularities in the complex s-plane, Cauchy's theorem leads to the finite energy sum rules (FESR) \cite{QCDSR2}, \cite{Shankar}-\cite{CIPT2}

\begin{equation}
 \int_{\mathrm{sth}}^{s_0} ds\; \frac{1}{\pi}\; p(s) \;Im \,\Pi(s)|_{HAD} \; = 
-  \frac{1}{2 \pi i} \; \oint_{C(|s_0|) }\, ds \;p(s) \;\Pi(s)|_{QCD} \;,
\end{equation}

where $p(s)$ is an analytic  weight kernel, $s_{th}$ is the hadronic threshold, and the finite radius of the circle, $s_0$, is large enough for QCD and the OPE to be used on the circle. Physical observables determined from FESR should be independent of $s_0$. In practice, though, this  is not exact, and there is usually a region of stability starting at $s_0 \gtrsim 2 - 4\, \mbox{GeV}^2$ in the light-quark sector where
observables are fairly independent of $s_0$. Equation (3) is the mathematical statement of what is usually referred to as quark-hadron duality. Since PQCD is not valid on the real axis in the time-like resonance region ($s \geq 0$), in principle there is a possibility of problems on the circle near the real axis, known as  duality violations (DV), an issue  identified very early in \cite{Shankar} long before the present formulation of QCDSR. In order to account for this potential issue it was first proposed in \cite{PINCH1}-\cite{PINCH2} to use suitable integration kernels pinched so that they vanish on the real axis. An underlying assumption in this approach is that QCD is still valid on the Cauchy circle provided the radius is large enough. This is a contentious issue, as there is an alternative proposal which relaxes this assumption and seeks suitable models to account for DV \cite{DV1}-\cite{DV2}. In any case, it should be kept in mind that DV effects are difficult to estimate as they are unknown by definition, as very clearly pointed out in \cite{DV1}.\\ 

Most of the early determinations of the vacuum condensates in the OPE from FESR, Eq.(1), were performed with simple kernels $p(s) = s^N$ and using the vector or axial-vector correlators together with data, e.g.  from $e^+ e^-$ annihilation in the light-quark sector, or $\tau$-lepton hadronic decays \cite{Eidelman}-\cite{SOLA}, as well as data on $e^+ e^-$ annihilation in the charm-quark region \cite{CHARM}-\cite{CHARMd}.  In the framework of fixed order perturbation theory \cite{FOPT} the FESR, Eq.(3), become

\begin{equation}
(-)^N C_{2N+2} \langle O_{2N+2}\rangle = \int_{0}^{s_0} ds\; s^N\;\frac{1}{\pi} \;Im \,\Pi(s)|_{HAD} \; 
-  \frac{s_0^{N+1}}{(N+1)}\; I_N(s_0)|_{PQCD}\;,
\end{equation}

where $N\geq 1$, and $I_N(s_0)|_{PQCD}$ is the integrated PQCD contribution. In this approach, and to next-to-leading order (NLO) in PQCD, radiative corrections to the condensates do not induce mixing of condensates of different dimension \cite{Launer}, a welcome feature. All of these early results relied on available PQCD information at the time, mostly only up to next-to-next-to leading order (NNLO), and on values of $\alpha_s$ considerably lower than at present, i.e. some 40\% lower. Due to this, the PQCD contribution to the FESR was a manageable correction leading to relatively high accuracy in the values of the condensates. This situation changed dramatically with the availability of radiative corrections at the five-loop level, and a considerably higher value of the strong quark-gluon coupling. As a result, current determinations based on Eq.(4) \cite{C2b}-\cite{GG}  are affected by such large uncertainties that the dimension $d=4$ gluon condensate is known with close to 100\% error, and no meaningful results are obtained for  condensates of higher dimension. For instance, the ALEPH Collaboration \cite{Davier} has used $\tau$-decay data \cite{ALEPH2} together with an indiscriminate global fit of all parameters, i.e. strong coupling and power corrections, to obtain an unphysical negative value for the gluon condensate. The source of the problem in this approach is the almost cancellation between two large and comparable quantities on the right hand side of Eq.(4). In other words,  large PQCD logarithmic terms tend to swamp the  power corrections in sum rules. Specifically, the condensates determined from FESR are the result of a difference between two integrals, one involving the data and the other PQCD on the circle of radius $s=|s_0|$. Both contributions are large and comparable, thus leading to a large uncertainty. An exception is the case of chiral condensates which can be determined with reasonable accuracy due to the absence of PQCD \cite{GG}, \cite{PINCH2}, \cite{DV2}, \cite{CHIRAL}.\\
The third approach to obtain the dimension $d=4$ power correction in  the OPE is based on QCDSR for the vector current correlator in the charm-quark region, where there is data from $e^+e^-$ annihilation into hadrons. Early determinations \cite{CHARM}-\cite{CHARMd} have been superseded due to the large increase of the strong coupling $\alpha_s$ over the years, and by the availability of NNLO perturbative information.\\

In this paper we discuss a novel determination of this condensate in the charm-quark region using the vector current correlator and involving a pinched integration kernel in the FESR exhibiting a singularity at the origin in the complex $s$-plane. This allows for (a) a substantial enhancement of the hadronic contribution due to the well known first two $\psi$-poles, followed by a large quenching of the resonance region above them, where the data has large uncertainties, and (b) an extraction of the gluon condensate entering in the Cauchy residue of the singularity at the origin through the low energy QCD expansion. This leads to an expression for the gluon condensate involving contributions from three terms, the experimental data, the high energy PQCD contribution and the low energy PQCD expansion in inverse powers of the heavy-quark mass. It turns out that the last two terms have opposite signs, thus rendering the total PQCD contribution to be one order of magnitude smaller than the data. This last feature circumvents the problem with traditional FESR where the condensates are the result of a fine balance between two large contributions, the hadronic and the PQCD integrals. Hence, this leads to a substantially more accurate result.\\

\section{Determination of $C_4 \langle O_4 \rangle$}
We  consider the vector current correlator

\begin{equation}
\Pi_{\mu\nu} (q^2) = i \int d^4x \; e^{iqx} \langle 0| T(V_\mu(x) \; V_\nu(0))|0\rangle 
= (q_\mu\; q_\nu - q^2 g_{\mu\nu})\; \Pi(q^2)\;,
\end{equation}

where $V_\mu(x) = \bar{c}(x) \gamma_\mu c(x)$. From Cauchy's residue theorem in the complex s-plane  one obtains

\begin{equation}
\int_{s_{th}=M^2_{J/\psi}}^{s_0}
p(s)\, \frac{1}{\pi} Im \,\Pi(s)\,ds = - \frac{1}{2\pi i}
\oint_{C(|s_0|)}
p(s) \,\Pi(s) \,ds 
 + \text{Res}[\Pi(s) \,p(s),s=0]\;,
\end{equation}

where $p(s)$ is now a meromorphic function, the integral on the right hand side  involves QCD, provided $s_0$ is large enough, and
the left hand side involves the hadronic spectral function 

\begin{equation}
Im\;\Pi(s) = \frac{1}{12 \pi} \;R_c(s) \;,
\end{equation}

with $R_c(s)$ the standard $R$-ratio for charm production in $e^+ e^-$ annihilation. Notice the lower limit of integration on the right hand side of Eq.(6). This threshold lies above the (suppressed) pure gluonic intermediate states entering at NNLO, thus not included in the observable $R_c$. It was found in \cite{Kuhn}  that the total background is different from $R_{uds}$ by $0.01\%$, and thus the non-$R_{uds}$ contributions are entirely negligible.

The PQCD piece of $\Pi(s)$, entering the  integral around the circle in Eq.(6), can be formally written as  

\begin{equation}
\Pi(s)|_{PQCD} = e_c^2 \;\sum_{n=0} \left( \frac{\alpha_s(\mu^2)}{\pi}\right)^n \; \Pi^{(n)}(s) \;,
\end{equation}

where $e_c = 2/3$ is the charm-quark electric charge, and 

\begin{equation}
\Pi^{(n)} (s) = \sum_{i=0} \left(\frac{\bar{m}_c^2}{s}\right)^i \; \Pi^{(n)}_i\;,
\end{equation}

with $\overline{m}_c \equiv \overline{m}_c(\mu)$  the running charm-quark mass in the $\overline{MS}$-scheme. Up to order $\cal{O}$ $[\alpha_s^2 (\bar{m}_c^2/s)^6]$ the function $\Pi(s)_{PQCD}$   has been calculated in \cite{QCD1},  exact results for $\Pi_0^{(3)}$ and $\Pi_1^{(3)}$ have been found in \cite{QCD2}, and  $\Pi_2^{(3)}$ is known  up to a constant \cite{QCD3}. At five-loop order, $\cal{O}$$(\alpha_s^4)$, the full logarithmic terms for $\Pi_0^{(4)}$ were determined  in \cite{QCD5}, and for $\Pi_1^{(4)}$ in \cite{QCD6}. Since there is incomplete knowledge at this order we shall use the available information as a measure of the truncation error in PQCD. There is also a non-perturbative QCD contribution to $\Pi(s)$, with the leading term being the gluon condensate. This contribution, though, is negligible on account of $s_0$ being large. However, the gluon condensate also  enters in the sum rules through the Cauchy residue in Eq.(6), provided $p(s)$ is singular at the origin, a feature that constitutes the essence of this determination.
The low energy expansion of the vector correlator around $s=0$  in PQCD can be written as

\begin{equation}
\Pi_{PQCD} (s) = \frac{3\, e_c^2}{16\, \pi^2}\; \sum_{n \geq 0} \overline{C}_n \; z^n\;,
\end{equation}

where $z = s/(4 \overline{m}_c^2)$. The coefficients $\overline{C}_n$ are then expanded in powers of $\alpha_s(\mu)$

\begin{eqnarray}
\bar{C}_n &=& \bar{C}_{n}^{(0)}+\frac{\alpha_{s}(\mu)}{\pi}\left(\bar{C}_{n}^{(10)}+\bar{C}_{n}^{(11)} l_{m}\right)
+\left(\frac{\alpha_{s}(\mu)}{\pi}\right)^2 \left(\bar{C}_{n}^{(20)}+\bar{C}_{n}^{(21)} l_{m}+\bar{C}_{n}^{(22)} l_{m}^{2}\right) \nonumber \\ [.3cm]
&+& \left(\frac{\alpha_{s}(\mu)}{\pi}\right)^3 \left(\bar{C}_{n}^{(30)}+\bar{C}_{n}^{(31)} l_{m}+\bar{C}_{n}^{(32)} l_{m}^{2}
+ \bar{C}_{n}^{(33)} l_{m}^{3}\right)+ \ldots
\end{eqnarray}

where $l_m\equiv \ln(\bar{m}_{c}^{2}(\mu)/\mu^2)$. Up to three loop level the coefficients of $\bar{C}_n$ are known  up to $n=30$  \cite{QCD8}-\cite{QCD9b}. At four-loop level  $\bar{C}_0$ and $\bar{C}_1$ were determined in  \cite{QCD8}-\cite{QCD8b}, \cite{QCD10}, $\bar{C}_2$ is from \cite{QCD9}-\cite{QCD9b}, and  $\bar{C}_3$ from \cite{QCD11}. We shall choose $p(s)$ so that no coefficients $\bar{C}_4$ and above contribute to the Cauchy residue at $s=0$. The different expansions in Eqs.(9) and (10) are to be understood as a result of the scale hierarchy $\Lambda_{QCD} << m_c << s_0$.
\begin{figure}
[ht]
\begin{center}
\includegraphics[height=2.5in, width=3.4in]{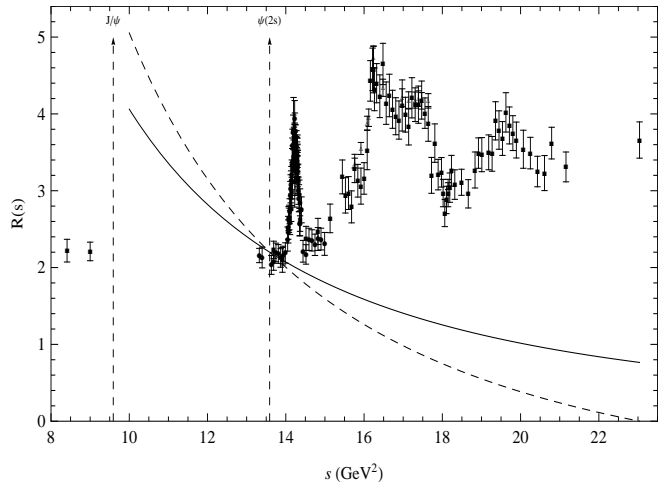}
\caption{\footnotesize{Experimental data for the total  $R(s)$ ratio \cite{EXP4}-\cite{EXP3} together with the optimal integration kernel, Eq.(17), with $N=2$ (dash curve), and $p(s) = 1/s^2$ (solid curve) normalized to coincide with the former at the position of the $\psi(2S)$ peak.}}
\end{center}
\end{figure}
The  non-perturbative contributions to the OPE  involve inverse  powers  of $q^2$, and the leading term, of dimension $d=4$, is the gluon condensate \cite{Broad} 

\begin{equation}
\lim_{-q^2 \rightarrow 0}\Pi(q^2)|_{NPQCD}(q^2) = - \frac{1}{q^4} \, \frac{\langle \frac{\alpha_s}{\pi} G^2 \rangle}{12 \pi} \left( 1 + {\cal{O}} (\alpha_s)\right).
\end{equation}

As is well known, in the heavy-quark sector there is no underlying chiral symmetry, and the heavy-quark condensate reduces to the gluon condensate, e.g. to leading order in $m_Q^{-1}$

\begin{equation}
\langle \bar{Q} Q \rangle = - \frac{1}{12 m_Q} \langle \frac{\alpha_s}{\pi} G^2\rangle \;.
\end{equation}

In the sequel we ignore potential renormalon ambiguities, as we are not aware of renormalon analyses in heavy-quark expansions, with masses expressed in the $\overline{MS}$-regularization scheme. Furthermore, in the present analysis we determine the dimension $d=4$ power correction in the OPE of the heavy-quark vector correlator. This term has traditionally been identified with the gluon condensate, and could also be viewed simply as a phenomenological parameter of the QCDSR approach. In other words, we are not determining the gluon condensate from first principles, as done e.g. in LQCD, which involves issues which may not arise in phenomenological extractions such as the one presented here.\\
 
Finally, the leading non-perturbative contribution to the FESR, Eq.(6), from singular kernels of the form $p(s) = 1/s^{N+1}$, with $N \geq 0$, has been calculated in \cite{Broad}. However, we shall make use of the result in \cite{Kuhn}, which is already expressed in the $\overline{MS}$-scheme, and to NLO reads

\begin{equation}
{\mbox{Res}} \left[\frac{\Pi(s)|_{NPQCD}}{s^{N+1}}, s=0\right]= \frac{e_c^2}{\left(4 \bar{m}_c^2\right)^{N+2}} \; \langle\frac{\alpha_s}{\pi} G^2\rangle 
\, a_N\, \left(1 + \frac{\alpha_s}{\pi} \, \bar{b}_N\right),
\end{equation}

where the quark mass and the coupling depend on $\mu$, and

\begin{equation}
a_N = - \frac{2 N + 2}{15} \, \frac{\Gamma(4 + N) \Gamma(7/2)}{\Gamma(7/2 + N) \Gamma(4)},
\end{equation}

\begin{equation}
\bar{b}_N = b_N - (2N + 4) \left( \frac{4}{3} - l_m\right) ,
\end{equation}

with $b_0 = 1469/162$, $b_1=135779/12960$, $b_2=1969/168$, and other values given in \cite{Kuhn},\cite{Broad}. The NNLO term is unknown so that we will include it as a source of uncertainty later. 
The fundamental QCD parameters are the charm-quark mass $m_c(\mu^2)$, the running strong coupling $\alpha_s(\mu^2)$, and the gluon condensate $\langle\frac{\alpha_s}{\pi} G^2 \rangle$. For the strong coupling we use the current value from lattice QCD (LQCD) \cite{LATTa} $\alpha_s(M_Z^2)= 0.1183 \pm 0.0007$, and the charm-quark mass also from LQCD \cite{LATTm} $\bar{m}_{c}({\mbox{3 GeV}}) = 986.4 \pm 4.1\;{\mbox{GeV}}$, which agrees with the most recent QCDSR  determination \cite{SBmc} $\bar{m}_{c}({\mbox{3 GeV}}) = 987 \pm 9\;{\mbox{MeV}}$. Solving the renormalization group equation for the strong coupling and for the quark mass one can obtain their values at any scale $s$ in terms of their values at any given reference scale, e.g. $s =s_0$ \cite{Davier}. Regarding the renormalization scale $\mu$, we follow the choice \cite{SBmc}-\cite{CHET} $\mu^2 = (3 \; {\mbox{GeV}})^2$ in the low energy QCD expansion, and $\mu^2 = s_0$ in the high energy QCD expansion on the circle of radius $s = |s_0|$.\\ 

\begin{table}
\begin{center}
\begin{tabular}{cccccccccc}
& \multicolumn{8}{c}{$\;\;\;\;\;\;\;\;\;\;\;\;\;\;\;\;\;\;\;\;$Uncertainties ($\mbox{GeV}^4$)}\\
\cline{3-9}
\noalign{\smallskip}
Method 	&	$\langle\frac{\alpha_s}{\pi} G^2 \rangle$	& $\Delta_{s_0}$& $\Delta_{\alpha_s}$ 	&  $\Delta_{ m_c}$ 	&		$\Delta_{ \text{DATA}}$ 	&	 $\Delta_{\text{T}}$ \\   
& (\small{$\mbox{GeV}^4$})&&&&&&&\\
\hline 
\noalign{\smallskip}
$(a)$									&			0.044		 																		&		0.0028							& 0.0003						& 0.0048									&  0.0043						& 0.007						\\

$(b)$									&			0.026		 																		&		0.0016								& 0.0001						& 0.0027									&  0.0024 	& 0.004								\\      
\hline
\end{tabular}
\caption{\footnotesize{Results for the gluon condensate for the kernel, Eq.(17), for $N=2$ and its sources of uncertainty  from the values of $s_0$, $\alpha_s$, $m_c$, the experimental data, and the total uncertainty. Method (a) refers to using the currently known NLO radiative correction to the residue, Eq.(14). Method (b) assumes that the NNLO correction is as large, and of the same sign as the NLO one (see text).}} 
\end{center}
\end{table}

Turning to the experimental data, we  follow closely the analysis of \cite{Kuhn},\cite{CHET_etal}. For the first two narrow resonances we use the latest data from the Particle Data Group \cite{PDG}, $M_{J/\psi}= 3.096916 (11)\; \mbox{GeV}$, $\Gamma_{J/\psi \rightarrow e^{+} e^{-}} = 5.55 (14) \;\mbox{keV}$, $M_{\psi(2s)}= 3.68609 (4)\; \mbox{GeV}$, $\Gamma_{\psi(2s) \rightarrow e^{+} e^{-}} = 2.35 (4) \;\mbox{keV}$. These  two narrow resonances are followed by the open charm region where it is necessary to subtract from the total $R$-ratio the contribution from the light quark sector, i.e. $R_{uds}$. We perform this subtraction as in \cite{MC2}. In the region $3.97\;\mbox{GeV} \leq \sqrt{s} \leq 4.26\;\mbox{GeV}$ we only use CLEO data \cite{EXP4} as they are the most precise. In connection with the two data sets from BES \cite{EXP2}-\cite{EXP3}, we assume  that the systematic uncertainties are not fully independent and add them linearly, rather than in quadrature. However, we treat these data as independent from the CLEO data set \cite{EXP4}, and thus add errors in quadrature. There is no data in the region $s = 25 - 49 \;\mbox{GeV}^2$, and   beyond  there is CLEO data up to $s\simeq 90 \;\mbox{GeV}^2$. The latter data is fully compatible with PQCD.\\

We discuss next the integration kernels $p(s)$ in Eq.(6), which we choose as

\begin{equation}
p(s) = \left(\frac{s_0}{s}\right)^N  - 1 \;,
\end{equation}

with $N \geq 2$. This choice is motivated by (i) the suppression of potential quark-hadron duality violations, as $p(s_0)=0$, and (ii) the simultaneous enhancement of the two ground state narrow resonances and the quenching of the resonance region contribution. This second feature can be appreciated from Fig. 1. In principle, the constant term in the kernel, Eq.(17), should not contribute to the sum rule, Eq.(6), due to the absence of a $d=2$ power correction. If quark-hadron duality were to be exact, then this would be an exact result. We find that
while numerically the line integral is not exactly equal to the integral around the circle, the contribution of this constant term in $p(s)$ to Eq.(6), i.e. the difference between the two integrals is small. However, we shall take this into account later in the final result.  Regarding the value of $N$, as discussed in \cite{Kuhn},\cite{CHET_etal}, inverse moments  $p(s) = 1/s^N$ should not involve too large values of $N$. In fact, the convergence of PQCD deteriorates with increasing $N$,  and the uncertainties in $\alpha_s$ and the renormalization scale $\mu$ have a greater impact on the total error of the result. We found that Eq.(17) with $N=2$ is the optimal kernel as explained next. In Fig.1 we show the experimental data for the ratio  $R(s)$ together with the kernel Eq.(17) with $N=2$ and for $s_0 \simeq \; 23 \; \mbox{GeV}^2$, and the simple kernel $p(s)= 1/s^2$ normalized such that both kernels coincide at the peak of the second narrow resonance $\psi(2S)$, i.e. $s \simeq 13.6\; \mbox{GeV}^2$. One can easily appreciate that in comparison with the latter, the former kernel leads to a welcome higher enhancement of the weight of the $J/\psi$ and the $\psi(2S)$, as well as to a stronger suppression of the broad resonance region,  particularly near the onset of the continuum. Also, the kernel, Eq.(17), with $N=2$ (i) leads to the most stable result for the gluon condensate as a function of $s_0$, and (ii) gives a result with the smallest uncertainty. In fact, varying $s_0$ from an initial value $s_0= 23.04 \,{\mbox{GeV}^2}$, corresponding to the last BES data point \cite{EXP2}-\cite{EXP3}, and $s_0= 30.0 \,{\mbox{GeV}^2}$ changes the value of the gluon condensate within the range determined by the uncertainties in $\alpha_s$ and $\bar{m}_c$
The contour integral evaluated using fixed order perturbation theory ($\mu^2=s_0$) gives essentially the same result as using contour improved perturbation theory.\\
In Table 1 we show the results, together with a breakdown of the relevant uncertainties due to the  various parameters.
The numerical value is $\langle  \frac{\alpha_s}{\pi} G^2 \rangle = 0.048 \,\pm\, 0.003 \, {\mbox{GeV}}^4$ from the kernel Eq.(17), and $\langle  \frac{\alpha_s}{\pi} G^2 \rangle = 0.041 \,\pm\, 0.003 \, {\mbox{GeV}}^4$ for $p(s) = 1/s^2$. Combining these results leads to $\langle  \frac{\alpha_s}{\pi} G^2 \rangle = 0.044 \,\pm\, 0.007 \, {\mbox{GeV}}^4$. Of some concern is the large size of the NLO radiative correction to the residue, Eq.(14), and the fact that the NNLO is unknown. Radiative corrections to condensates at NNLO are currently known only for the quark condensate entering the Adler function \cite{radcond}, and it is of the same sign as the NLO term. Adopting the conservative procedure of assuming the NNLO to be of the same size and sign as the NLO gives $\langle  \frac{\alpha_s}{\pi} G^2 \rangle = 0.026 \,\pm\, 0.002 \, {\mbox{GeV}}^4$. Including this uncertainty into the gluon condensate gives our preferred value

\begin{equation}
\langle  \frac{\alpha_s}{\pi} G^2 \rangle = 0.037 \,\pm\, 0.015 \, {\mbox{GeV}}^4 \;,
\end{equation}

This result for the gluon condensate agrees within errors with a recent LQCD value \cite{LATTa} $\langle  \frac{\alpha_s}{\pi} G^2 \rangle = 0.028 \,\pm\, 0.003 \, {\mbox{GeV}}^4$. Another LQCD determination \cite{LATTb} reports a still smaller value  consistent with zero $\langle  \frac{\alpha_s}{\pi} G^2 \rangle = 0.002 \,\pm\, 0.002 \, {\mbox{GeV}}^4$. 
On the other hand, our result is   larger than our most recent value from the corrected ALEPH data base \cite{GG} which, however, has a very large uncertainty, i.e. $\langle  \frac{\alpha_s}{\pi} G^2 \rangle = 0.005 \,\pm\, 0.004 \, {\mbox{GeV}}^4$. As mentioned earlier, such a large uncertainty in the traditional FESR method is due to the condensate resulting from the difference between two large integrals involving  PQCD and the data.
Very early determinations from QCDSR in the heavy-quark sector \cite{QCDSR1}-\cite{QCDSR2}, \cite{CHARM}-\cite{CHARMd} can be summarized in the value

\begin{equation}
\langle  \frac{\alpha_s}{\pi} G^2 \rangle = 0.018 \,\pm\, 0.012 \, {\mbox{GeV}}^4 \;.
\end{equation}

A comparison with our result, Eq.(18), is not straightforward mainly because (i) our method differs substantially from others as it requires not only high energy QCD information but also the low energy QCD expansion. Both contributions to the gluon condensate are comparable but of different sign, thus becoming an order of magnitude smaller than the data contribution, a more than welcome feature. And (ii) current PQCD information at high energy is far more detailed than 20-30 years ago, and the value of $\alpha_s$ is currently much higher.  A more recent QCDSR value in the light-quark region, from an unconventional method, gives \cite{rho}
\begin{equation}
\langle  \frac{\alpha_s}{\pi} G^2 \rangle = 0.062 \,\pm\, 0.019 \, {\mbox{GeV}}^4 \;,
\end{equation}
in agreement within errors with our value, Eq.(18). 
The result above would support the view that the gluon condensate is channel/sector independent \cite{QCDSR1}-\cite{QCDSR2}, \cite{ZAKHA}. 

\section{Conclusion}
In this paper we have introduced a novel approach to determine the dimension $d=4$ power correction to the OPE, traditionally identified with the gluon condensate. The method relies on QCD FESR, but it is not based on the standard FESR, which involve the difference between two large quantities, i.e. the PQCD integral around the Cauchy circle in the complex $s$-plane and the line integral of the data along the real and positive $s$-axis. Instead, we considered FESR involving a suitable integration kernel, singular at the origin in the $s$-plane, which (i) invites the gluon condensate to enter the FESR in a leading role through the Cauchy residue in Eq.(6), and (ii) in the hadronic sector it enhances substantially the contribution of the well known narrow resonances, while strongly quenching  the region beyond. Feature (i) results in the gluon condensate being determined by the data, and by both the low and the high energy QCD expansions of the vector correlator. The latter two are of opposite sign, leading to a partial cancellation with a total value close to one order of magnitude smaller than the contribution from the data. Hence, this feature avoids the shortcomings of the standard FESR approach, where there is only one (large) PQCD contribution of similar size as the data contribution.
The impact of uncertainties in all relevant parameters entering this determination was assessed, and shown in Table 1. A relevant source of, perhaps, the larger systematic uncertainty is the lack of knowledge of NNLO radiative correction to the gluon condensate. This enters the Cauchy residue, Eq.(14). We attempted to account for this issue by assuming that the NNLO radiative correction is as large as the NLO one. Our final result is compatible with some LQCD values, and previous QCDSR results. By confronting it with results from the light-quark sector it supports the widely accepted view that the gluon condensate is channel/sector independent \cite{QCDSR1}-\cite{QCDSR2},\cite{ZAKHA}.  \\

\begin {Large}
{\bf Acknowledgements}
\end{Large}
 The authors thank Sebastian Bodenstein for discussions. This work was supported in part by the National Research Foundation (South Africa), and by the Deutsches Forschungsgemeinschaft (Germany).

\end{document}